# Magnetic wire-based sensors for the µ-rheology of complex fluids

L. Chevry[1], N. K. Sampathkumar[1], A. Cebers[2] and J.-F. Berret[1*]

[1] *Matière et Systèmes Complexes, UMR 7057 CNRS Université Denis Diderot Paris-VII, Bâtiment Condorcet 10 rue Alice Domon et Léonie Duquet, F-75205 Paris, France*
[2] *Department of Theoretical Physics, University of Latvia, Zellu 8, Riga LV-1002, Latvia*

**Abstract:** We propose a simple µ-rheology technique to evaluate the viscoelastic properties of complex fluids. The method is based on the use of magnetic wires of a few microns in length submitted to a rotational magnetic field. In this work, the method is implemented on a surfactant wormlike micellar solution that behaves as an ideal Maxwell fluid. With increasing frequency, the wires undergo a transition between a steady and a hindered rotation regime. The study shows that the average rotational velocity and the amplitudes of the oscillations obey scaling laws with well-defined exponents. From a comparison between model predictions and experiments, the rheological parameters of the fluid are determined.

# 1 - Introduction

Rheology is the study of flow and deformation of fluids when they are submitted to mechanical stresses. Conventional rheometers determine the relationship between strain and stress in steady or oscillating flow on samples of a few milliliters. µ-rheology in contrast studies the motion of micron-size probe particles that are thermally fluctuating *via* the interactions with a surrounding medium, or particles that are forced by an external field. In the first case, the µ-rheology is said to be passive, in the second active. In both cases, the motion of the probes is related by the mechanical properties of the medium. Fluids produced in small quantities, e.g. costly protein dispersion or fluids confined in small volumes down to 1 picoliter, such as living cells can only be examined by this technique.

With the development of microfluidics systems in the last decade [1], rapid advances were made in the field of µ-rheology. Standard experimental protocols and data treatment softwares are now available and implemented on a regular and controlled basis [2-7]. The correspondence between µ- and macro-rheology is nowadays well established. In µ-rheology, the objective is to translate the motion of a probe particle into the relevant rheological quantities of the fluid such as the elastic complex modulus or the creep response function. These quantities are expressed as a function of the frequency or of the time [8]. With passive motion, this transformation is done using the Generalized Stokes Einstein equation that relates the mean-squared displacement of the probe and the fluid parameters [4-7, 9, 10]. With active motion, it is obtained by comparing the force applied to the probe and the distances by which it moved [4, 11, 12]. The conversion of the time or frequency dependences of some local µ-rheology variables into a macroscopic rheological response remains however a challenging issue. Several factors can impact this conversion, such as the signal-to-noise ratio, the variability of the probe or of its environment, the bandwidth in frequency, the time scale etc. In this context, methods able to retrieve the fluid parameters directly, *i.e.* without passing through complex data treatments are highly desirable. The derivation of such a method is the first objective of the present study.

As a second objective, we aim to test anisotropic colloids in active µ-rheology conditions, and to determine to what extent the probes are effective for viscoelastic measurements. Most µ-rheology techniques developed so far are taking advantage of nano- or micron-size spherical beads. These methods are both powerful and quantitative. Several studies, however, reported the use of anisotropic objects such as disks [13], rods [14-20] and wires [21-23], for both passive [13, 21, 23] and active [14, 16-18, 22, 24-28] µ-rheology. Some recent work has shown that rod or wire based µ-rheology experiments could also bring significant advances to the field [18-20, 23, 27]. Plasmonic nanorod absorbers were for instance developed as orientation sensors using coupled planar illumination microscopy imaging [19] or polarization-sensitive photothermal imaging [20]. Each of these methods provides the particle trajectory and orientation as a function of the time, an important outcome for the study of anisotropic or heterogeneous media. For driven motions, the transition between steady and hindered rotation of a micro-actuator above a critical frequency was successfully used to evaluate the viscosity of the fluid [18, 24, 26-28]. In the later case, only Newton liquids were surveyed.

Here we study the rotational properties of magnetic wires and examine their potential as probe particles for active µ-rheology. As emphasized in recent reviews [29, 30], wires at the micrometer range represent an active research topic owing to their unique applications in mesoscopic physics and fabrication of nanoscale devices. With respect to µ-rheology, wires exhibit also distinct advantages as compared to beads. Submitted to a magnetic field, wires are behaving more like a conventional stress rheometer, in the sense that a torque is applied and results in the rotation of the probe. Moreover, due to their large aspect ratio, wires can probe a medium over different length scales, and be sensitive to its heterogeneities. In the present work, we show that by monitoring the motion of wires under a steady rotational field, the viscoelastic parameters of a Maxwell fluid can be determined accurately. Two physical quantities describing the motion of the wire were introduced and measured as a function of the frequency: the average rotation velocity and the amplitude of the oscillation in the instable regime. These quantities display characteristic asymptotic behaviors as a function of the frequency and bear some similarities with those of the elastic complex modulus of the fluid.

# 2 – Theoretical Background

The torque exerted on a magnetic nanowire suspended in a liquid and submitted to a constant magnetic excitation $\vec{H}$ expresses as:

$$\vec{\Gamma}_M = \mu_0 V \vec{m} \wedge \vec{H} \qquad (1)$$

where $V$ is the volume of the wire ($V = \frac{\pi}{4}D^2 L$), $\vec{m}$ its magnetization and $\mu_0$ the permeability in vacuum. For a superparamagnetic wire of susceptibility $\chi$ [18], Eq.1 becomes:

$$\Gamma_M = \frac{1}{2}\mu_0 V \Delta\chi H^2 \sin(2\beta) \qquad (2)$$

Here, $\Delta\chi = \chi^2/(2+\chi)$ and $\beta$ is the angle between the wire and the orientation of the applied field, as shown in Fig. 1.

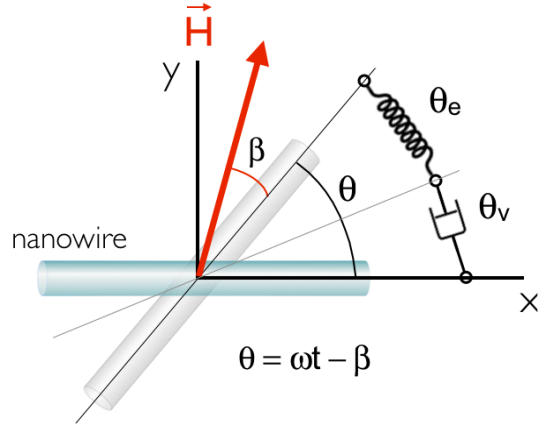

*Figure 1*: *Schematic representation of a wire submitted to a steady rotational magnetic excitation $H(t)$. Immersed in a viscoelastic Maxwell fluid, a wire experiences two restoring torques noted $\Gamma_V$ and $\Gamma_E$, associated to the rotation angles $\theta_V$ and $\theta_E$ respectively (Eqs. 3 and 4). In the model, the Maxwell fluid is described as a spring and a dashpot in series. $\theta$ denotes the wire orientation and $\beta$ the retardation angle with respect to the field.*

Under the application of a magnetic torque $\Gamma_M$, the wire rotates in a propeller-like motion so as to minimize the angle $\beta$ and to eventually align with $\vec{H}$. With a field rotating at the frequency $\omega$, one has the relation $\beta = \omega t - \theta$, where $\theta$ describes the wire orientation (Fig. 1). Immersed in a viscoelastic Maxwell fluid, a wire experiences two restoring torques that slow down its rotation. One torque has a viscous origin and writes:

$$\Gamma_V = \frac{\pi \eta_0 L^3}{3g\left(\frac{L}{D}\right)}\frac{d\theta_V}{dt} \qquad (3)$$

and one torque has an elastic origin and writes:

$$\Gamma_E = \frac{\pi G_0 L^3}{3g\left(\frac{L}{D}\right)}\theta_E \qquad (4)$$

In Eqs. (3-4), $\eta_0$ is the static viscosity of the fluid, $G_0$ its elastic modulus and $g\left(\frac{L}{D}\right)$ is a dimensionless function of the anisotropy ratio $p = L/D$. In this study, we assume that $g(p) = \ln(p) - 0.662 + 0.917\,p - 0.050\,p^2$ which is valid in the interval $2 < p < 20$ [31]. In rheology, the Maxwell model is depicted as a spring and a dashpot in series. In this configuration, the elastic and viscous deformations are additive, and the shear stresses are equal. By analogy, we assume here that $\theta = \theta_V + \theta_E$, and that $\Gamma_M = \Gamma_V = \Gamma_E$, resulting in the differential equation:

$$\frac{d\theta(t)}{dt}(1 + \theta_0 \cos 2(\omega t - \theta)) = \omega_C \sin 2(\omega t - \theta) + \omega \theta_0 \cos 2(\omega t - \theta) \quad (5)$$

where

$$\omega_C = \frac{3}{8}\frac{\mu_0 \Delta \chi}{\eta_0} g\left(\frac{L}{D}\right)\frac{D^2}{L^2}H^2 \quad (6)$$

and

$$\theta_0 = \frac{3}{4}\frac{\mu_0 \Delta \chi}{G_0} g\left(\frac{L}{D}\right)\frac{D^2}{L^2}H^2 \quad (7)$$

In Eqs. 6 and 7, the two quantities $\omega_C$ and $\theta_0$ vary quadratically with the magnetic excitation and $\theta_0 = 2\omega_C \tau_R$. Here, $\tau_R$ denotes the characteristic relaxation time of the fluid ($\tau_R = \eta_0/G_0$). For the data treatment, the geometrical characteristics of the wire can be advantageously combined into a single dimensionless parameter $L^* = L/D\sqrt{g(L/D)}$. In these conditions, Eq. 6 becomes:

$$\omega_C = \frac{3\mu_0 \Delta \chi}{8\eta_0}\frac{H^2}{L^{*2}} \quad (8)$$

In recent studies, the case of the Newton fluid was evaluated [18, 24, 26-28]. It was shown that for frequency above $\omega_C$ the wire undergoes an hydrodynamic instability between two rotation regimes [32, 33]. In Regime I, the wire rotates at the same frequency as the field, whereas in Regime II it is animated of asynchronous back-and-forth motions. Because wires are superparamagnetic, the frequency of the oscillations far form the instability ($\omega \gg \omega_C$) is twice that of the excitation. The present study extends these prior observations to a fluid that is viscoelastic. Fig. 2 illustrates the rotational phase diagram derived from the resolution of Eq. 5. The phase behavior of a wire immersed in a Maxwell fluid is described using a 3-dimensional representation where the variables are $\omega_C, \omega$ and $\theta_0$. In this representation, the planes $(\omega_C, \omega)$ and $(\omega, \theta_0)$ describes the behavior of a purely viscous fluid and that of a purely elastic solid, respectively. The phase diagram of Fig. 2 displays again two main regimes of synchronous (Regime I, $\omega \leq \omega_C$) and asynchronous (Regime II, $\omega > \omega_C$) rotations. Interestingly, the back-and-forth instability limit found for the Newtonian case persists for non-zero elasticity ($\theta_0 \neq 0$). The panels **a** to **f** in Fig. 2 illustrate typical temporal evolutions of the wire orientation for viscous (**a**, **b**), elastic (**c**, **d**) and viscoelastic (**e**, **f**) cases. The red lines indicate the average angular velocity $\Omega = \overline{d\theta/dt}$ [18, 32, 33]. For viscous and viscoelastic fluids, $\Omega$ is positive whereas it is zero for an elastic solid. For $\theta_0 \geq 1$, the rotation of the wire displays a second instability that manifests itself by an abrupt back motion after a period of increase (Fig. 2d and 2f). The overall behavior in this range remains however similar to that found for $\theta_0 < 1$, with a periodic back-and-forth motion and a positive average angular velocity.

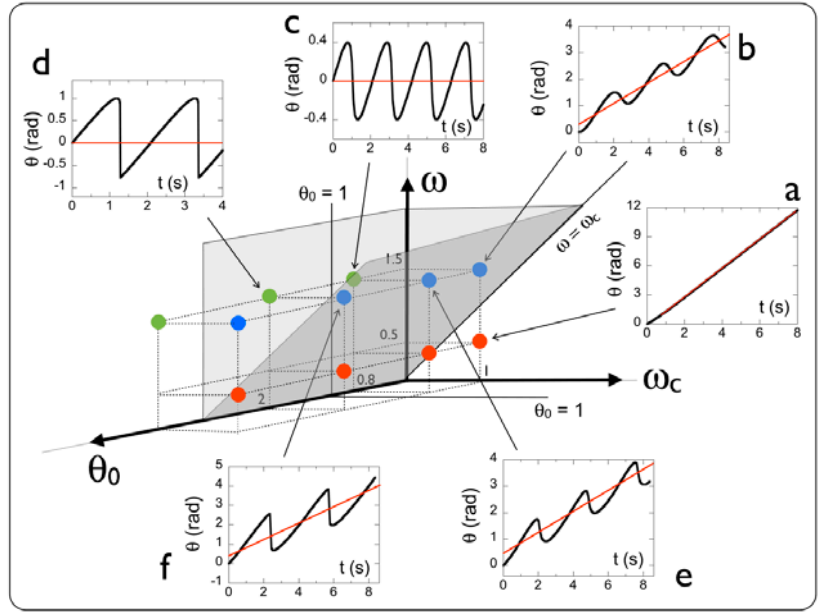

*Figure 2:* Rotation phase behavior of a wire immersed in a Maxwell fluid using a 3-dimensional representation $(\omega_C, \omega, \theta_0)$. Here $\omega$ is the angular frequency of the rotating field, $\omega_C$ the angular frequency of the hydrodynamic instability between the synchronous and asynchronous regimes, and $\theta_0 = 2\omega_C \tau_R$. In this representation, the planes $(\omega_C, \omega)$ and $(\omega, \theta_0)$ describes the behaviors of a purely viscous fluid and of a purely elastic solid, respectively. The insets **a-f** illustrate the time dependence of the wire orientation angle $\theta$ in the different regimes. The red straight lines in **a-f** represent the average angular velocity $\Omega = \overline{d\theta/dt}$ in each condition.

## 3 - Materials and Methods

*3.1 - Magnetic wires*

Wires were formed by electrostatic complexation between oppositely charged nanoparticles and polymers [34, 35]. The particles were 10.7 nm iron oxide nanocrystals ($\gamma$-$Fe_2O_3$, maghemite) synthesized by polycondensation of metallic salts in alkaline aqueous media [36]. An extensive characterization of the nanometric $\gamma$-$Fe_2O_3$ using various techniques, including vibrating sample magnetometry, dynamic light scattering, zetametry and transmission electron microscopy is provided in Supporting Information (S1 – S2). To improve their colloidal stability, the cationic particles were coated with $M_W = 2100 \: g \: mol^{-1}$ poly(sodium acrylate) (Aldrich) using the precipitation-redispersion process [37]. This process resulted in the adsorption of a highly resilient 3 nm polymer layer surrounding the particles. The co-assembly process leading to the formation of stiff magnetic wires followed a bottom-up approach, where the elementary bricks were the iron oxide nanoparticles and the "gluing" agent a highly charged polycation [35, 38], here poly(diallyldimethylammonium chloride) (Aldrich) of molecular weigth $M_W < 100000 \text{ g mol}^{-1}$. Fig. S3 illustrates the desalting protocol designed for the elaboration of the magnetic wires and displays wires observed by optical microscopy (Fig. S3a) and by transmission electron microscopy (Fig. S3b). For the µ-rheology experiments, the wires were characterized by an average length $L_0 = 20 \: \mu m$ and a polydispersity of 0.5. For this sample, the lengths were comprised between 1 and 50 µm.

Electrophoretic mobility and ζ–potential measurements made with a Zetasizer Nano ZS Malvern Instrument showed that the wires were electrically neutral [35]. The shelf life of the co-assembled structures is of the order of several years.

*3.2 - Wormlike micellar solutions and linear rheology*

The surfactant solutions investigated here were mixtures of cetylpyridinium chloride ($CP^+; Cl^-$) and sodium salicylate ($Na^+; Sal^-$) (abbreviated as CPCl/NaSal) dispersed in a 0.5 M NaCl brine [39, 40]. Since the pioneering work of Rehage and Hoffman [41], CPCl/NaSal is known to self-assemble spontaneously into micrometer long wormlike micelles. The surfactant solution was prepared at $c = 2\,wt.\%$. Under these conditions, the micelles build a semi-dilute entangled network that imparts to the solution its Maxwell viscoelasticity. In the semi-dilute regime, the mesh size of the network is of the order of 30 nm, *i.e.* much smaller than the wire diameter. The frequency dependence of the elastic complex modulus $G^*(\omega) = G'(\omega) + iG''(\omega)$ was obtained on a CSL 100 rheometer (TA Instruments) using a cone-and-plate and controlled shear rate. Dynamical measurements were carried out for angular frequency $\omega = 0.1 - 100$ rad s$^{-1}$ at temperature $T = 27\,°C$. The viscoelastic response of CPCl/NaSal wormlike micelles was that of a Maxwell fluid with a unique relaxation time $\tau_R$. The viscoelastic parameters $G_0$, $\tau_R$ and $\eta_0 = \lim_{\omega \to 0} |G^*(\omega)|/\omega$ were derived from dynamical frequency sweeps.

*3.3 - Micro-rheology*

Bright field and phase-contrast microscopy was used to monitor the actuation of the wires as a function of time. Stacks of images were acquired on an IX71 inverted microscope (Olympus) equipped with 100× objectives. For optical microscopy, 35 µl of a dispersion containing the wires were deposited on a glass plate and sealed into to a Gene Frame® (Abgene/Advanced Biotech) dual adhesive system. The glass plate was introduced into a homemade device generating both static and rotational magnetic fields, thanks to two pairs of coils working with a 90°-phase shift. An electronic set-up allowed measurements in the frequency range 1 mHz - 100 Hz and at magnetic fields B = 0 – 20 mTesla. A stream of nitrogen directed toward the measuring cell was used to thermalize the sample at the desired temperature. The image acquisition system consisted of an EXi Blue CCD camera (QImaging) working with Metaview (Universal Imaging Inc.). Images of wires were digitized and treated by the ImageJ software and plugins (http://rsbweb.nih.gov/ij/).

# 4 - Results and discussion

4.1 – Determination of the magnetic susceptibility

To determine the magnetic properties of the wires *i.e.* the susceptibility parameter $\Delta\chi$ in Eq. 2, rotation experiments were carried out on a viscous fluid of known viscosity. A 85 wt. % glycerol-water mixture of static viscosity $\eta_0 = 0.044$ Pa s$^{-1}$ (T = 32°C [23]) served as a suspending medium for the $\gamma$-Fe$_2$O$_3$ wires. In this experiment, the typical length and diameter of the wires were in the range of 5 to 20 µm and of 0.5 to 1 µm respectively. The critical frequency $\omega_C$ between the synchronous (Regime I) and asynchronous (Regime II) regimes was measured on 65 different wires submitted to a magnetic field of 10.4 mT. In Fig. 3, $\omega_C$ is plotted as a function of the dimensionless parameter $L^* = L/D\sqrt{g(L/D)}$. The critical frequency was found to decrease as $\omega_C \sim L^{*-2}$, in excellent agreement with the prediction of Eq. 8. From the prefactor ($3\mu_0 \Delta\chi H^2/8\eta_0 = 1800 \pm 400$ rad s$^{-1}$), we infer $\Delta\chi = 2.1 \pm 0.4$, and $\chi = 3.4 \pm 0.4$. The wire susceptibility is here 5 times larger than that obtained in our first report [34]. This increase is attributed to the increase of size of the γ-Fe$_2$O$_3$ particles used in the synthesis, 10.7 nm instead of 7.0 nm in Ref. [34]. Knowing $\Delta\chi$, the magnetic torque applied to a wire is evaluated quantitatively.

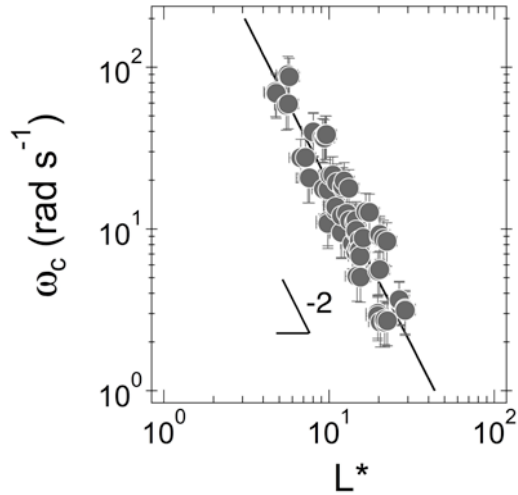

*Figure 3:* *Critical frequency $\omega_C$ as a function of the parameter $L^* = L/D\sqrt{g(L/D)}$ obtained for wires dispersed in a 85 wt. % glycerol-water mixture of static viscosity $\eta_0 = 0.044$ Pa s$^{-1}$ (T = 32°C [23]). The straight line is calculated using Eq. 8 and $\Delta\chi = \chi^2/(2 + \chi) = 2.1 \pm 0.4$. Here, L, D and $\chi$ are the length, diameter and susceptibility of the wire, respectively.*

4.2 – Micro-rheology of wormlike micelles

Macro- and microrheology experiments were performed on a $c = 2\, wt.\%$ CPCl/NaSal micellar solution at $T = 27°C$. In the Supporting Information (S4), the elastic and loss moduli, $G'(\omega)$ and $G''(\omega)$, and the complex viscosity $\eta(\omega)$ obtained by cone-and-plate rheometry are shown at this temperature. With increasing frequency, $G'(\omega)$ increases quadratically and reaches a plateau, whereas $G''(\omega)$ exhibits a resonance-like profile peaked at $\omega = 1/\tau_R$. The data were adjusted with the Maxwell model using the expressions $G'(\omega) = G_0 \omega^2 \tau_R^2/(1 + \omega^2 \tau_R^2)$ and $G''(\omega) = G_0 \omega \tau_R/(1 + \omega^2 \tau_R^2)$ for the elastic and loss moduli, respectively. The complex viscosity modulus was adjusted using $\eta(\omega) = \eta_0/\sqrt{1 + \omega^2 \tau_R^2}$, where $\eta_0 = G_0 \tau_R$ denotes the static viscosity. At T = 27 °C, the viscoelastic parameters are $G_0 = 7.1 \pm 0.1$ Pa, $\tau_R = 0.14 \pm 0.01$ s and $\eta_0 = 1.0 \pm 0.1$ Pa s. These results confirm those published for the first time on the same system two decades ago [39, 40, 42].

In Fig. 4, a rotating magnetic field of 10.4 mT was applied to a 8.1 µm nanowire immersed in the CPCl/NaSal solution at increasing frequencies, between 0.1 and 20 rad s$^{-1}$ (inset). Although the wires were assembled from charged particles and polymers, we did not observe any degradation of the structures when they were dispersed in the viscoelastic fluid. The motion of the wires was monitored by optical microscopy, and the time dependence of their orientation was derived. Figs. 4a-d show 4 time traces $\theta(t)$ obtained at $\omega = 0.14$, 0.40, 2.9 and 17.0 rad s$^{-1}$, respectively. At low frequency, the wire rotates in phase with the field, and $\theta(t) = \omega t$. Above the critical frequency (Eq. 6), here $\omega_C = 0.38$ rad s$^{-1}$, the wires are animated of back-and-forth motion characteristic of the asynchronous regime (Regime II) and $\theta(t)$ displays oscillations. As illustrated in Figs. 4b-4d, the frequency of the back-and-forth increases with that of the magnetic excitation.

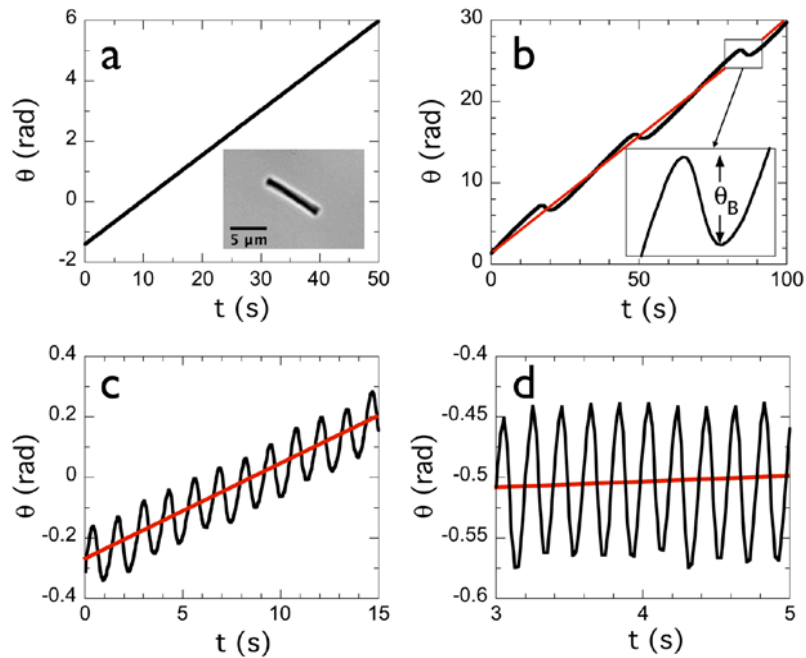

***Figure 4:*** *Rotation angle $\theta(t)$ of a 8.1 µm wire as a function of the time at various frequencies: a) $\omega = 0.14$ rad s$^{-1}$; b) $\omega = 0.40$ rad s$^{-1}$; c) $\omega = 2.9$ rad s$^{-1}$ and d) $\omega = 17.0$ rad s$^{-1}$. Inset in a): image of the wire by bright field microscopy (100×). Inset in b): $\theta_B$ is defined as the angle by which the wire returns back after a period of increase. The red straight lines in **a-d** represent the average angular velocity $\Omega = \overline{d\theta/dt}$.*

To interpret the data, we define two quantities that characterize the behavior of the wire: the average angular velocity $\Omega = \overline{d\theta/dt}$ given by the slope of the straight lines in Fig. 4, and the angle $\theta_B$ by which the wire returns back after a period of increase. $\theta_B$ corresponds to the region where $d\theta/dt < 0$ (inset in Fig. 4b). In the next section, the dependencies of $\Omega(\omega)$ and of $\theta_B(\omega)$ are first discussed.

### 4.3 – Correspondence between macro- and micro-rheology

Fig. 5 displays the evolution of the average angular velocity normalized by the critical frequency, $\widetilde{\Omega} = \Omega/\omega_C$ as a function of the reduced frequency $X = \omega/\omega_C$ obtained for 5 nanowires ($L = 8 - 14 \mu m$) in different conditions of magnetic field and rotation frequency. With increasing frequency, the average velocity increases, passes through a cusp-like maximum at the critical frequency ($X = 1$), and then decreases. The data in Fig. 5 were adjusted using the stationary solutions of Eq. 5 [32]:

$$X \leq 1 \qquad \widetilde{\Omega}(X) = X$$
$$X \geq 1 \qquad \widetilde{\Omega}(X) = X - \sqrt{X^2 - 1} \qquad (9)$$

Interestingly, the set of equations in Eq. 9 is identical to that of a Newton fluid, indicating that elasticity does not play a role in the onset of the instability, or on the dependence of the average velocity with frequency [18]. From the $\omega_C$-values obtained, the viscosity of the fluid was calculated and averaged over the different measurements. It was found to be $\eta_0 = 1.3 \pm 0.3$ Pa s, in excellent agreement with the cone-and-plate rotational rheometry value at this temperature, $\eta_0 = 1.0 \pm 0.1$ Pa s. Details on the determination of the viscoelastic parameters using µ-rheology can be found in S5.

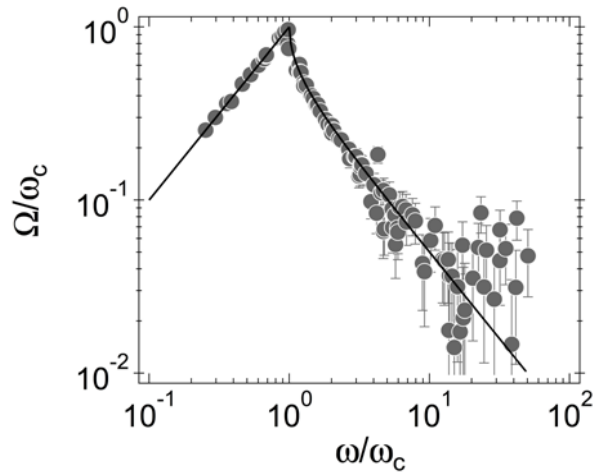

*Figure 5*: *Average angular velocity Ω divided by the critical frequency $\omega_C$ as a function of the reduced frequency $\omega/\omega_C$ for nanowires of length $L = 8 - 14$ µm in different conditions of magnetic field and rotation frequency (S5). The solid line corresponds to the best fit using the stationary solutions of Eq. 5.*

Fig. 6 displays the frequency dependences of $\theta_B(\omega)$, the angle by which the wire returns after a period of increase, for wires of different lengths, $L = 8.2, 6.3$ and $8.1$ µm (from blue to green). $\theta_B$ being related to the asynchronous regime, it is not defined for $\omega \leq \omega_C$. For frequencies slightly above $\omega_C$, the angle decreases with increasing frequency according to a power law of the form $\theta_B(\omega) \sim \omega^{-\alpha}$ (where $\alpha$ is close to unity), and then it flattens into a frequency independent plateau. The height of the plateau depends on the experimental conditions, and in particular on the critical frequency $\omega_C$. The transition between the $\theta_B(\omega) \sim \omega^{-\alpha}$ dependence and the plateau occurs at $\omega \approx 3$ rad s$^{-1}$ in the present case. The continuous lines between the data are calculated according to Eq. 5 using $\tau_R$ as a unique adjustable parameter. For $\omega\tau_R \gg 1$, Eq. 5 also predicts that $\lim_{\omega\to\infty} \theta_B(\omega) = \theta_0 = 2\omega_C\tau_R$. The limiting angles $\theta_0$ obtained in the different configurations are indicated on the figure by arrows. The inset in Fig. 6 illustrates the $\theta_B(\omega\tau_R)$-behaviors for Newton and Maxwell fluids on a broader frequency range. They are calculated according to Eq. 5 using $\omega_C = 0.1$ rad s$^{-1}$ and $\tau_R = 1$ s. For the Newton fluid, the amplitude of the oscillations decreases with increasing frequency. In contrast, the Maxwell fluid displays a crossover between the viscous and elastic regimes and it occurs at a fixed value of the reduced frequency, here $\omega\tau_R = 1/2$.

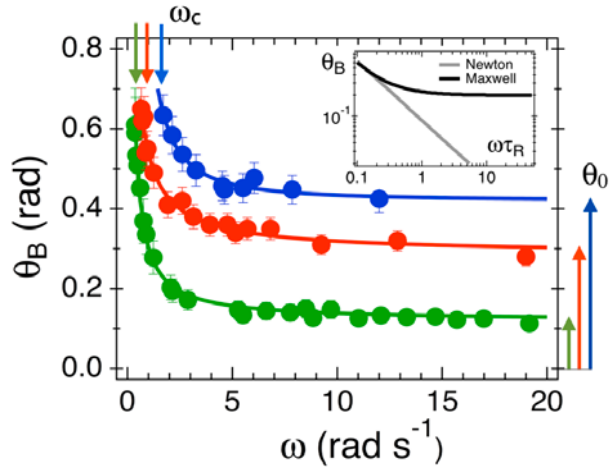

*Figure 6*: *Frequency dependence of the angle $\theta_B(\omega)$ obtained for wires of different lengths (L = 8.2, 6.3 and 8.1 µm from blue to green) and for B = 10.4 mT. At frequencies such as $\omega\tau_R \gg 1$, $\theta_B(\omega)$ exhibits a plateau of height $\theta_0 = 2\omega_C\tau_R$. The continuous lines result from best fits using the Maxwell model and Eqs. 5-7. Inset: $\theta_B(\omega\tau_R)$-behaviors for Newton and Maxwell fluids calculated from Eq. 5 using $\omega_C = 0.1$ rad s$^{-1}$ and $\tau_R = 1$ s.*

|  | $\eta_0$ (Pa s) | $G_0$ (Pa) | $\tau_R$ (s) |
|---|---|---|---|
| cone-and-plate rheology | 1.0 ± 0.1 | 7.1 ± 0.1 | 0.14 ± 0.02 |
| wire-based µ-rheology technique | 1.3 ± 0.3 | 9.4 ± 2.0 | 0.14 ± 0.02 |

*Table I*: *Viscoelastic parameters $\eta_0, G_0$ and $\tau_R$ of a CPCl/NaSal wormlike micellar solution at concentration $c = 2$ wt.% and temperature $T = 27\,°C$. The results for the wire-based µ-rheology technique are averaged over 9 independent measurements using 9 different wires (see Supporting Information S5).*

The value of $1/2$ is explained by the fact that the frequency of the oscillations in Regime II is twice that of the field. The existence of a plateau in the amplitude of the oscillations at high frequency equivalent to that of the pure elastic solid (Fig. 2c) represents a strong evidence of the elastic character of the fluid. From the plateau region at $\theta_0 = 2\omega_C\tau_R$, the relaxation time $\tau_R$ characteristic of the micellar solution was determined, and found in good agreement with the macro-rheology measurements. For the wire-based method, one obtained $\tau_R = 0.14 \pm$ 0.03 s, which compares well with $\tau_R = 0.14 \pm 0.01$ s derived from cone-and-plate data. The elastic modulus $G_0$ is then calculated from the viscosity and from the relaxation times. One gets $G_0 = 9.4 \pm 2$ Pa, again in agreement with macro-rheology ($G_0 = 7.1 \pm 0.1$ Pa). The viscoelastic parameters obtained by the two techniques are compared in Table I.

To confirm the adequacy of the model, the time traces of rotating wires were adjusted using the solutions of Eq. 5. Fig. 7a-d shows the rotation angle data (closed symbols) of a 8.1 μm wire at $\omega$ = 0.4, 1.2, 2.9 and 7.8 rad s$^{-1}$ as a function of the product $\omega t$. The adjustment was performed using $\omega_C$ and $\theta_0$ as fitting parameters, and provided excellent results (continuous curves in red). For this set of data, we found $\omega_C = 0.38$ rad s$^{-1}$ and $\theta_0 = 0.12$ rad, corresponding a static viscosity $\eta_0 = 1.4$ Pa s and to a modulus $G_0 = 9.2$ Pa. These later values are again in good agreement with those of the cone-and-plate rheology (Table I).

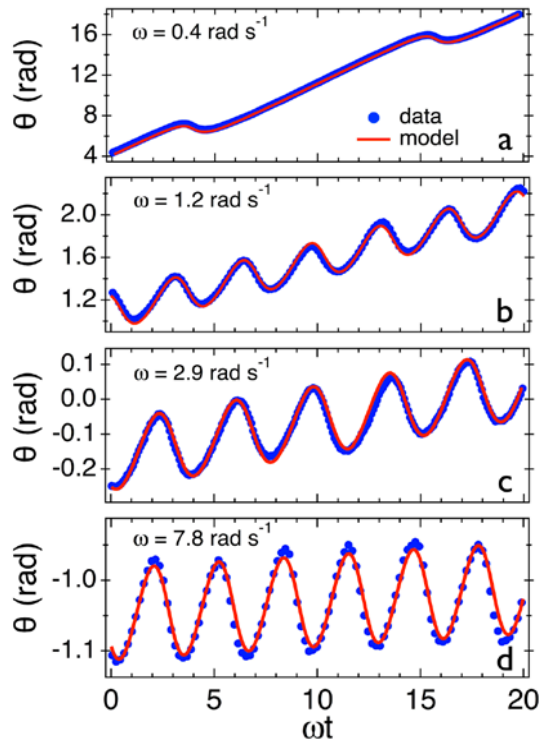

*Figure 7*: *Comparison between experimental (closed symbols) and predicted (continuous lines) behavior of a 8.1 μm wire submitted to a 10.4 mT magnetic field at frequencies ω = 0.4 (a), 1.2 (b), 2.9 (c) and 7.8 (d) rad s$^{-1}$. The angle of rotation is shown as a function of the product ωt. For the fitting, $\omega_C$ (= 0.38 rad s$^{-1}$) and $\theta_0$ (= 0.12 rad) in Eq. 5 were treated as adjustable parameters.*

In the Introduction, it was mentioned that the conversion of local μ-rheology variables into macroscopic parameters may be difficult or subjected to large uncertainties [11, 43] and that alternate approaches need to be considered. Our experiments on wires submitted to rotational magnetic field show that such approaches exist and can be realized. For that, we defined two physical quantities describing the motion of the wires, $\Omega(\omega)$ and $\theta_B(\omega)$. The study reveals a formal analogy between the frequency dependences of the two quantities and those of the complex modulus. As $G'(\omega)$ and $G''(\omega)$ for a Maxwell fluid, $\Omega(\omega)$ and $\theta_B(\omega)$ exhibit power law dependences as a function of $\omega$. From the asymptotic behaviors at low and high frequency the fluid parameters can be derived. The correspondence is the most noticeable for $G'(\omega)$ and $\theta_B(\omega)$: both display a transition between a viscous and an elastic regime, and both exhibit a frequency independent plateau at high frequency. Differences between these two sets of variables exist however. $\Omega(\omega)$ and $G''(\omega)$ do show maxima, but their positions on the frequency axis are different, $\omega_C$ for $\Omega(\omega)$ and $1/\tau_R$ for $G''(\omega)$.

# 5 - Conclusion

In this paper, we demonstrate that micron-size wires with magnetic properties can serve as probe particles for active µ-rheology experiments. Submitted to a rotating magnetic field with increasing angular frequency, the behavior of the wires immersed in viscous and in viscoelastic fluids is modeled. The rotational phase diagram exhibits a transition between a synchronous and an asynchronous regime, this second regime being characterized by a back-and-forth motion [18, 24, 26-28]. The instability resembles that observed on laboratory benches when a viscous solution is actuated with a magnetic bar. The difference with the present work is in the size of the bar, and also in its magnetic property. As demonstrated by the scaling in Fig. 3, the wires are superparamagnetic and not ferromagnetic as magnetic bars are. Hence, for a given value of the magnetic field the torque applied is known with accuracy (Eq. 2). For the viscoelastic fluid, the transition occurs at a critical frequency $\omega_C$ that is independent of the elastic modulus $G_0$. For the elastic solid, similar calculations are performed and show that the wires are oscillating around their initial position with a constant amplitude. These predictions are tested using nanostructured magnetic wires of length 5 - 20 µm dispersed in a cetylpyridinium chloride/sodium salicylate solution of wormlike micelles. The most significant findings to emerge from this work are i) the excellent agreement between the model and the experiments performed, and ii) the accurate derivation of the viscoelastic parameters $\eta_0$, $G_0$ and $\tau_R$ of the fluid. These findings have also important implications. They demonstrate *a posteriori* that the experiments with micron-size wires are indeed consistent with the linear regime of shear deformation [44]. Here, we do not observe nonlinear effects in the frequency dependences of the average velocity $\Omega(\omega)$ and the angle $\theta_B(\omega)$. These two last quantities are found to display well defined scaling behaviors apart from $\omega\tau_R = 1/2$. The static viscosity is derived from $\Omega(\omega)$, whereas the relaxation time and the modulus are obtained from $\theta_B(\omega)$.

Tested on a Maxwell fluid for sake of simplicity, the wire-based technique can be extended in principle to any kind of fluids. For fluids characterized by a distribution of relaxation times, it is possible to define a Generalized Maxwell Model describing the rotation of a wire. The model is based on a differential equation of the kind of Eq. 5. In its concept, the approach is similar to that developed for the elastic and loss moduli $G'(\omega)$ and $G''(\omega)$, where each mode of the distribution contribute to the overall viscoelastic response. Interestingly, even in the case of a fluid with several relaxation modes, the average rotation frequency $\Omega(\omega)$ will still exhibit a transition behavior similar to that of Fig. 3, allowing the determination of the static viscosity. If the fluid is gel-like and characterized by an infinite static viscosity, the wire responses will be comparable to those illustrated in Figs. 2 for systems in the $\omega_C = 0$ plane. This wire-based technique is also especially adapted for time dependent effects and kinetics, since any temporal change in the viscoelasticity will result in a change of the average velocity and of the oscillations amplitude. In conclusion, we have developed an easy and straightforward technique to measure the linear rheological properties of complex fluids in confined environment.

# Acknowledgments

Discussions with Bérengère Abou, Rémy Colin, Marc-Antoine Fardin, Jérôme Fresnais, Sandra Lerouge are acknowledged. The Laboratoire Physico-chimie des Electrolytes, Colloïdes et Sciences Analytiques (UMR Université Pierre et Marie Curie-CNRS n° 7612) is acknowledged for providing us with the magnetic nanoparticles.

# Supporting Information

The Supporting Information section S1 provides the characterization of the nanoparticle in terms of sizes and distribution. S2 deals with the vibrating sample magnetometry results obtained with iron oxide nanoparticle dispersions. The desalting transition that allows to produce the nanostructured wires is described in S3. S4 provides the macrorheology characterization of the wormlike micellar fluid. S5 lists the characteristics of the wires investigated.